\documentclass[%
 reprint,
 amsmath,amssymb,
aip,
]{revtex4-1}

\usepackage{graphicx}
\usepackage{dcolumn}
\usepackage{bm}
\usepackage[T1]{fontenc}
\usepackage[utf8x]{inputenc}
\usepackage{tikz}
\usepackage[version=4]{mhchem} 
\usepackage{adjustbox}
\usepackage{multirow}
\usepackage{threeparttable}
\usepackage{gensymb}

\newcommand*\wnumb{cm$^{-1}$}
\newcommand*\sorb[1][]{\textit{s\textsuperscript{\hspace{0.03cm}#1}}}
\newcommand*\porb[1][]{\textit{p\textsuperscript{\hspace{0.03cm}#1}}}
\newcommand*\dorb[1][]{\textit{d\textsuperscript{\hspace{0.03cm}#1}}}
\newcommand*\forb[1][]{\textit{f\textsuperscript{\hspace{0.03cm}#1}}}
\newcommand{\Lower}[1]{\smash{\lower 1.5ex \hbox{#1}}}

\begin{document}

\preprint{APS/123-QED}

\title{Changing the paradigm in f-containing cold molecules: the impact of spin-orbit coupling and f-d transitions on quasi-bound vibrational states. 
}

\author{Marta Gałyńska}
\author{Matheus M. F. de Moraes}%
\email{matheusmorat@gmail.com}
\author{Paweł Tecmer}
\email{ptecmer@fizyka.umk.pl}
\author{Katharina Boguslawski}
\email{k.boguslawski@fizyka.umk.pl}

\affiliation{Institute of Physics, Faculty of Physics, Astronomy, and Informatics, Nicolaus Copernicus University in Toru\'{n}, Grudzi{a}dzka 5, 87-100 Toru\'{n}, Poland}
 
\date{\today}

\begin{abstract}
Present-day state-of-the-art ab initio many-body calculations on f-block containing cold molecules heavily focus on perturbative approaches for spin-orbit coupling and exclude a substantial part of the atomic transitions in the \forb{}- and \dorb{}-shell.
Here, we demonstrate the cruciality of a proper relativistic treatment of the \forb{}- and \dorb{}-shell in Yb-containing diatomics and the inclusion of $f\rightarrow d$ transitions to obtain physically sound elastic scatterings and pre-dissociation lifetimes.
We focus on state-of-the-art relativistic many-body calculations for the Yb atom's ground- and excited-state and the YbLi$^+$ potential energy surface.
For that purpose, we exploit various quantum many-body methods, namely a spin-free and four-component implementation of the coupled cluster singles and doubles (CCSD) model and its equation of motion extensions, spin-free complete active space self-consistent field, and internally contracted multi-reference (MR) configuration interaction approaches, and spin-free MRCCSD with a perturbative and full triples correction. 
We oppose scalar relativistic calculations to four-component variants to support the reliability of our EOM-CCSD study and shed new light on the interplay between these systems' spin-orbit coupling and the proper treatment of relativistic effects. 
Most importantly, we observe a significant shift in the electronic spectra of the $f\rightarrow d$ excitation block. We also provide new reference potential energy surfaces for ground and excited states for which theoretically sound elastic scattering and pre-dissociation lifetimes are calculated.

\end{abstract}

\maketitle

%
%
\section{Introduction}\label{sec:introduction}
The recent discovery of laser cooling~\cite{shuman-first-laser-cooling-diatomic-nature-2010} rendered experimental investigations on compounds in their electronic ground state at temperatures close to absolute zero possible.
During the last two decades, many sophisticated experimental methods such as helium buffer gas cooling,~\cite{hutzler-buffer-gas-cooling-cr-2012} photoassociation,~\cite{ulmanis-photoassociation-cr-2012} and magnetic Feshbach resonances\cite{bauer-control-magnetic-feschbach-mp-2009} were extensively applied to explore the chemistry and physics of so-called cold and ultracold atoms and molecules, which are typically studied in the range of 1.0 K and 1.0 mK, respectively.
These experiments facilitate measurements of atoms or molecules with remarkable precision by controlling their specific quantum state.
Quantum many-body methods, ~\cite{bartlett_2007, monika_mrcc, lischka2018multireference} on the other hand, can predict all fundamental properties associated with the ground and lowest-lying excited states of atoms and molecules at absolute zero temperature. 
Thus, cold and ultracold chemistry can combine theory and experiment to discover fundamental properties of matter in the quantum realm. 

However, experiments on cold and ultracold molecules are not trivial. 
Problematic or difficult experimental techniques are, for example, laser cooling for molecular systems (due to their complex hyperfine structure arising from the interaction between the magnetic and electric field of the nuclei and electrons as well as their rovibronic coupling) and evaporative cooling (due to the large probability of three-body loss collisions). 
At the same time, temperatures close to absolute zero are fairly suitable for quantum mechanical simulations as the thermal motion is minimized and the systems are predominantly in their ground state configurations.
Highly accurate quantum many-body calculations can provide much sought-after local properties of cold molecules. 
Examples are bond lengths,~\cite{kosicki-srf-caf-jpca-2017, visentin-yb2-pes-reexamination-2021} scattering constants,~\cite{borkowski-scattering-lengths-rbyb-pra-2013, kedziera-scattering-pra-2015} electric dipole moments,~\cite{tomza-ybli-pra-2015} electronic and vibrational spectra,~\cite{sorensen-rbyb-jpca-2009, tohme-liyb-jcp-2015, pawel-yb2, gomes-ybf-pccp-2021} which on their part provide fundamental insights on the interaction of atoms and molecules.~\cite{pawlak-np-2017, tomza-rmp-2019}
Furthermore, theoretical data on cold molecules is of the utmost importance in experimental studies to, for instance, parameterize experimental setups, define starting points for laser-induced methods, and interpret or validate experimental results.

In the early days, the cold chemistry community focused on alkali atoms because of their simple hydrogen-like structure and alkali metal diatomic systems. 
Thereafter, ytterbium-based molecules became highly valuable. 
Ytterbium's closed f-shell and the 4f${^1}$${^4}$6s${^2}$ ground-state electronic configuration make the electronic structure of the ytterbium atom similar to the group-II atoms. 
The simplest example is the \ce{YbLi+} system.
Tomza \textit{et al.}~\cite{tomza-ybli-pra-2015} conducted a theoretical study on \ce{YbLi+} focusing on the Yb$^+$ and Li atoms as precursors.
That study obtained and analyzed four singlet and five triplet states.
The closed-shell nature of the system allowed for ground-state computations using the canonical coupled-cluster singles and doubles and perturbative treatment of triple excitations (CCSD(T)) from the equilibrium region up to the dissociation limit.
Excited states were modeled using the  CCSD(T) and equation of motion coupled cluster singles and doubles with perturbative triples (EOM-CC3) methods.
Although the investigated electronic spectrum of \ce{YbLi+} covers the range of 25000--35000~\wnumb{} at the equilibrium and dissociation regions, respectively, no state correlated to the Yb(4f$^{13}$5d6s$^2$) electronic configuration was reported.
Studies on similar diatomic ytterbium-containing molecules suggest that there is more than one state to consider (other than f$^{14}$) and that these states' position strongly depends on the degree of correlation included.~\cite{dolg-ybh-ybf-cp-1992, su-ybx-uccsdt-cpl-2009}
Very recently, Pototschnig \textit{et al.}~\cite{gomes-ybf-pccp-2021} showed for YbF that some Yb$^+$(4f$^{13}$6s) correlated states have $T_e$ values around 5000~\wnumb{} and effectively cross the ground state correlated with the Yb$^+$(4f$^{14}$6s) electronic configuration.
Therefore, these states must be considered in quantum many-body calculations to avoid potentially non-negligible effects on the ground and low-lying states of \ce{YbLi+}. 
Unfortunately, standard multi-reference methods, like the complete active space self-consistent field~\cite{roos_casscf} (CASSCF) theory or the multi-reference configuration interaction (MRCI) approach, are not flexible enough to cope with many open-shell 4\forb[], 6\sorb[], 6\porb[], and 6\dorb[] configurations on an equal footing. 

Closed-shell electronic structures, like \ce{YbLi+}, are difficult to cool and trap experimentally.
Consequently, the experimental focus has been shifted towards open-shell doublet and triplet state molecules.
The YbX (X = Li, Na, K, and Rb) doublet ground-state series is a well-known example.~\cite{sorensen-rbyb-jpca-2009, borkowski-scattering-lengths-rbyb-pra-2013, tohme-liyb-jcp-2015, tohme-nayb-cpl-2015, tohme-kyb-ctc-2016}
Similar to the closed-shell \ce{YbLi+} molecule, theoretical calculations for \ce{YbLi} unveiled low-lying states correlated with the Yb(4f$^{13}$5d6s$^2$) electronic configuration. 
Considering the possible presence of both \forb[13] and \forb[14] configurations of Ytterbium, reliable theoretical studies have to provide a balanced description of both electronic configurations.

Lu and Peterson~\cite{lu-and-petterson-cc-dk3-basis-lanthanides-jcp-2016} showed for a lanthanide series that the spin-orbit coupling (SOC) effect on the ionization potentials from the 6$s$ and $5d$ shell are substantially smaller than removing an electron from the $4f$-shell.
Specifically, for the $4f$-shell ionization (Yb$^{2+}$(4\forb[14])$\to$ Yb$^{3+}$(4\forb[13]) the SOC value increases to 4~400~\wnumb{}.
Therefore, a reliable theoretical picture of the ionized and excited-state electronic structures must first correctly model all configurations relevant to the problem.
This includes, but is not restricted to, 4\forb{} excitations.
Then, SOC has to be included as it can drastically alter the relative energy among states.~\cite{cao-dolg-2003-ip-lanthanides-actinides-mp-2003}

This work aims to remedy this problem by providing a balanced treatment of relativistic and electronic correlation effects for the ground and excited-state electronic structures of \ce{YbLi+} by using the four-component (4C) Dirac--Coulomb Hamiltonian and various coupled-cluster methods.
Our study shed new light on the electronic structures and spectroscopic characterization of the Yb atom and the  \ce{YbLi+} molecule, providing a new reference point for experimental manipulations.  
%
%
\section{Computational details}\label{sec:comput-details}
\subsection{4C relativistic calculations\label{sec:soc-comput-details}}
The all-electron relativistic calculations were done using the 4C Dirac--Coulomb Hamiltonian, in which the $\left(SS|SS \right )$ integrals were approximated by a point charge model,~\cite{lvcorr} in the~\texttt{Dirac19} software package.~\cite{dirac-program-jcp-2020}
All computations utilized the all-electron dyall.v3z basis sets.~\cite{DyLight, gomes-la-lu-basis-tca-2010}  
The ground-state electronic calculations were computed with the relativistic version of the CCSD(T) method.~\cite{rel-ccsd-ijqc-95, rel-ccsd-t-jcp-1996} 
The low-lying excited states of the closed-shell Yb atom and the \ce{YbLi+} molecule were investigated using the relativistic version of the equation of motion coupled cluster singles and doubles (EOM-CCSD) method.~\cite{4c-eom-jcp-2018}
For the Yb atom, we correlated spinors in the energy range of [-2; 100 000] $E_{\text h}$.
For the \ce{YbLi+} molecule, spinors in the energy range of [-4; 100] $E_{\text h}$ were correlated in the CC calculations. 
That corresponds to 12 occupied and 253 virtual spinors in the Yb atom, and 13 occupied and 203 virtual spinors in the \ce{YbLi+} molecule. 
Among others, these include the occupied Yb:5\porb[], Yb:4\forb[], and Yb:6\sorb[], and virtual Li:2\sorb[], Li:2\porb[], Li:3\sorb[],  Yb:6\porb[], Yb:5\dorb[], Yb:7\sorb[], and Yb:6\dorb[] orbitals. 

\subsection{Spin-free and perturbative spin-orbit calculations}\label{sec:sf-comput-details}
The spin-free CCSD(T) and EOM-CCSD calculations were performed with the spin-free Dyall Hamiltonian~\cite{dyall-spin-free-hamiltonian} and the same computational setup as described in subsection~\ref{sec:soc-comput-details}.

Internally contracted multireference configuration interaction~\cite{molpro-internally-contracted-mrcisd-ee-tca-1992} (MRCI) calculations with the Davidson correction~\cite{davidson+q} (MRCI+Q) were performed using the \texttt{Molpro2019} software package.~\cite{molpro2020-authors, molpro-wires}
Internally contracted multi-reference coupled cluster (MRCCSD, MRCCSD(T), and MRCCSDT) calculations were carried out in the \textit{General Contracted Code} (GeCCo) software package.~\cite{gecco_2011, gecco_2012}
The all-electron cc-pVnZ-DK3 (n=T, Q) basis set~\cite{lu-and-petterson-cc-dk3-basis-lanthanides-jcp-2016} was utilized for Yb and cc-pVnZ-DK (n=T, Q) basis sets~\cite{prascher2011gaussian} was employed for the Li atom.
The scalar relativistic effects were accounted for using the third-order Douglas--Kroll--Hess (DKH3) approach.~\cite{wolf_2002, wolf2002generalized, reiher_2004a, reiher_2004b, reiher_book, dkh_book}

All multireference calculations used Restricted Active Space (RAS) orbitals to generate the reference wave function.
The active space is composed of 17 electrons distributed in 16 orbitals, RAS(17,16), and covers Yb:4\forb{}, Yb:5\dorb{}, and Yb:6\sorb{}, Li:2\sorb{}, and Li:2\porb{} orbitals.
All calculations used the C$_{2v}$ Abelian point group symmetry. 
Specifically, the active space contains six orbitals in symmetry $A_1$, four in $B_1$/$B_2$, and two in $A_2$, denoted as (6,4,4,2). 
In the restricted orbital occupation calculations, the active space was divided into RAS1(2,2,2,1) and RAS2(4,2,2,1), containing the Yb:4\forb{}-shell, and the remaining inner-valence orbitals, respectively.
Two sets of calculations were carried out to obtain the occupation-specific 4\forb[13] and 4\forb[14] states.
The molecular orbitals were optimized in a state-average fashion.
For the occupation-specific 4\forb[14] set of calculations, 43 states were averaged, including 22 $A_1$, 12 $B_1/B_2$, and 9 $A_2$ states.
For the 4\forb[13] set, 41 states were considered: 15 $A_1$, 14 $B_1/B_2$, and 12 $A_2$ states.

A similar procedure to \ce{YbLi+} was employed for multireference calculations on Yb and \ce{Yb+}.
The 4\forb[14] active space included RAS2(4,2,2,1), encompassing Yb:6\sorb{}, Yb:6\porb{}, and Yb:5\dorb{} orbitals.
For the 4\forb[13] states, the RAS1(2,2,2,1) active space was used, comprising Yb:4\forb{} orbitals and RAS2(4,2,2,1) comprising Yb:6\sorb{}, Yb:6\porb{}, and Yb:5\dorb{} orbitals.
The core-valence correlation was also considered for the Yb:5\sorb{} and Yb:5\porb{} orbitals.

The SOC was included perturbatively on top of spin-free energies by diagonalizing the matrix of electronic and spin-orbit operators. Using the one- and two-electron Breit--Pauli operators\cite{soc-mrci} in the basis of the spin-free eigen states ($\Lambda$+S) electronic of the Breit--Pauli Hamiltonian.
\subsection{Vibrational analysis}
We used the QuAC software~\cite{aoto_quac24} to perform a rovibrational and scattering analysis of the computed PESs.
Rovibrational parameters were obtained by numerical integration of the radial Schr\"odinger equation
\begin{equation}
    \Bigg[\frac{\partial^2}{\partial r^2}-\frac{J(J+1)}{r^2}+ V(r)-E_{v,J}\Bigg]\chi_{v,J}=0,
\end{equation}
where $J$ ($J\in \mathbb{N}$) is the rotational quantum number and $V(r)$ is the fitted electronic potential using a cubic spline expansion.
The resulting rovibrational levels were then fitted to obtain the Dunham parameters,~\cite{dunham-expansion-pr-1932} 
\begin{align}
    E(v,J)=&\omega_e(v+1/2)-\omega_ex_e(v+1/2)^2+ B_eJ(J+1).
\end{align}
In the above equation, $\omega_e$ is the vibrational constant, $\omega_ex_e$ accounts for the anharmonicity in the vibrational energy levels, and $B_e$ denotes the rotational constant. 
The phase-shift parameter $\eta_J(E)$ has been computed via the method of partial waves.~\cite{Bransden-blue-book}
The derivative of the phase-shift with respect with the collisional energy has used to compute numerically the collisional time-delay $\tau_J$(E) via the five-point method.\cite{leroyJCP71_54_5114}
%
%
\section{Results}

\begin{table*}[t]
\caption{The excitation energies and ionization potentials of the Yb atom calculated with wave function-based methods including icMRCISD, icMRCISD+Q, and 4c EOM-CCSD. The experimental values are obtained from Ref.\citenum{meggers_yb_atom}.}
\begin{tabular}{ccccccc}\hline
 Main &
Term & J & MRCISD&MRCISD+Q&4c EOM&Exp.~\citenum{meggers_yb_atom}\\ 
 config. & & & & & -CCSD &\\
\hline
\multicolumn{7}{c}{Excitation energies} \\
\hline
\multirow{3}{*}{4f$^{14}$6s6p}&\multirow{3}{*}{$^3$P$_u$}&0 &15 349&16 186 & 17 091& 17 288\\
                     &       &1 &16 103&16 932 & 17 800& 17 992\\
                     &       &2 &18 054&18 891 & 19 567& 19 710 \\
\cline{1-7} 
  \multirow{3}{*}{4f$^{14}$6s5d}&\multirow{3}{*}{$^3$D$_g$}& 1 &25 487&25 319 &25 659&24 489\\
                     &       & 2 &25 756&25 595 &25 881& 24 751 \\
                     &       & 3 &26 263&26 095 &26 271& 25 270 \\
\cline{1-7} 
  4f$^{14}$6s6p       &$^1$P$_u$ & 1 &26 685 &26 931 &26 159&25 068\\
\cline{1-7} 
  \multirow{4}{*}{4f$^{13}$5d6s$^2$}  &\multirow{4}{*}{$\big(\frac{7}{2}, \frac{3}{2}\big)_u$}& 2 &24 134&28 928 &15 601 &23 188\\
                    &                                      & 5 &26 661&31 502  &19 124&25 859 \\
                    &                                      & 3 &28 202&33 035  &21 043&27 445 \\
                    &                                      & 4 &29 012&33 786  &22 012&28 184 \\
\cline{1-7} 
 \multirow{6}{*}{4f$^{13}$5d6s$^2$}  &\multirow{6}{*}{$\big(\frac{7}{2}, \frac{5}{2}\big)_u$}& 6 &28 285&33 173 &20 500   &27 315\\
&   & 2 &29 288&34 038 &21 701&28 196 \\
 &&  1 &31 181&35 978  &20 911&28 857 \\
 &&  4 &30 957&35 701  &23 699&29 775\\ 
 &&  3 &31 310&36 107  &24 136&30 207 \\
 &&  5 &31 693&36 493  &24 565&30 525 \\
\cline{1-7} 
  \multirow{6}{*}{4f$^{13}$5d6s$^2$}&\multirow{6}{*}{$\big(\frac{5}{2}, \frac{5}{2}\big)_u$}& 0 &35 698&40 481 &27 518& n/a\\
 && 1 &39 079&43 951 &31 676& n/a\\
 && 5 &39 373&44 207 &32 023& n/a\\
 && 2 &40 249&45 064 &33 082& n/a\\
 && 3 &41 520&46 301 &34 549& n/a\\
 && 4 &42 017&46 775 &35 167& n/a \\
\cline{1-7} 
  \multirow{4}{*}{4f$^{13}$5d6s$^2$}&\multirow{4}{*}{$\big(\frac{5}{2}, \frac{3}{2}\big)_u$}& 4 &36 780&41 640 &29 542& n/a\\
 && 2 &37 894&42 645 &30 802& n/a\\
 && 1 &39 259&44 451 &34 590& n/a\\
 && 3 &40 136&44 918 &33 653& n/a\\
\cline{1-7}
 4f$^{14}$6s5d      &$^1$D$_g$ & 2 &28 424&28 795&28 259&27 678\\
\hline
\multicolumn{7}{c}{Ionization potentials}\\
 \hline
4f$^{14}$6s        &$^2$S$_g$ &1/2&47 944&48 791&50 478&50 443\\
\cline{1-7}
 \multirow{2}{*}{4f$^{13}$6s$^2$}&\multirow{2}{*}{$^2$F$_u$}&7/2&58 882&63 157&67 600&71 862\\
                    &          &5/2&69 206&73 481&78 252&82 011\\
\hline

\end{tabular}
\label{tab:Yb_atom}
\end{table*}

\subsection{Yb atom}
We will start with the assessment of our theoretical models in describing the electronic structures of the closed-shell Yb and open-shell \ce{Yb+} atoms and their spectra.
We will compare our results against the experimental data presented in Refs.~\citenum{meggers_yb_atom, meggers_yb_ion}.
Specifically, we exploit various levels of relativistic Hamiltonians and electronic structures methods.
Our results are collected in Table~\ref{tab:Yb_atom}.
The electronic excitations can be further divided into 4\forb[14] or 4\forb[13] states.
We should stress that the computed SO splittings agree with the experimental data, irrespective of the chosen configuration (4\forb[14] or 4\forb[13]) and methodology (perturbative SOC or 4c Hamiltonian).
On the other hand, the accuracy of the relative energies widely varies with the employed methodology and the state's configuration.

The first excited state of the Yb atom is the \forb[14] state $^3$P$_u$(6s~$\rightarrow$~6p), where the electron is excited from the Yb $^1$S$_g$ ground state, followed by the $^3$D$_g$(6s~$\rightarrow$~5d) state, which split into three $\Omega$ states.\cite{nist}
These low-lying \forb[14] states ($^3$P$_u$ and $^3$D$_g$) as well as the first ionization potential ($^2$S$_g$) calculated using 4c SOC-EOM-CCSD are similar to the SOC-MRCISD+Q results.
Furthermore, 4c SOC-EOM-CCSD almost exactly reproduces the experimental values (differences are between 200 and 1200 \wnumb{}) .
The error for the ionization potential of the Yb$^+$ \forb[13]6\sorb[2] state reduces with the level of theory according to SS-RASSCF > MRCISD > MRCISD+Q > MRCCSD > EOM-CCSD $\approx$ MRCCSD(T) $\approx$ MRCCSDT (with errors of -35~000, -13~000, -9~000, -6~500 and -4~000~\wnumb{}).
The SS-RASSCF error highlights that dynamic correlation considerably differs in the 4\forb[14] and the 4\forb[13] states.
Such a gap in the correlation energy is similar to the one reported for the third and fourth ionization potentials.~\cite{cao-dolg-2003-ip-lanthanides-actinides-mp-2003}
Thus, the main factor for this gap emerges from the change in the number of the 4\forb{} electrons and not in the system's charge.
The similar performance of MRCCSD(T) and MRCCSDT supports this case.
Hence, we can conclude that the 4\forb[14] correlations are crucial and need to be properly accounted for by including either higher excitations or sufficiently large virtual spaces.
Finally, the difference in error between MRCCSD and EOM-CCSD gives us an estimate of the error caused by the lack of orbital relaxation in the reference of the latter, which is around 2~500~\wnumb{}.

For the neutral atom, the targeted \forb[13] states include the excitations from the Yb 4\forb[] to the 5\dorb[] shell, resulting in the Yb 4\forb[13]6\sorb[2]5\dorb[] configuration.
For these \forb[13] states, the 4c SOC-EOM-CCSD errors in the Yb 4\forb[] to 5\dorb[] excitations increase to approximately -6~000 and -8~000~\wnumb{} compared to the \forb[14] states.
Similarly, the error of the MRCISD+Q results increases to around 5~000~\wnumb{}.
We should stress that the error decreases to below 2~000~\wnumb{} without the Davidson correction.
This behavior indicates a cancellation of errors at the MRCISD level.
Most likely, the error from the 4$f^{13}$-5$d^1$ interaction cancels out the one from the 4$f^{14}$ shell at the MRCISD level, while the lack of orbital relaxation leads to a higher absolute error in EOM-CCSD calculations.

Consequently, the relative errors in (excitation and ionization) energies compared to the experimental value appear to primarily originate from the missing 4\forb[14] correlation energy, either due to the lack of higher excitations or the limited virtual space.
This difficulty might be directly transferable to ytterbium-containing molecules.
Therefore, we should be aware of a similar underestimation in relative energies among states correlated to these 4$f^{13}$ configurations in the YbLi$^+$ cation.

\subsection{YbLi$^+$ Electronic Structure}

\begin{figure*}
\includegraphics{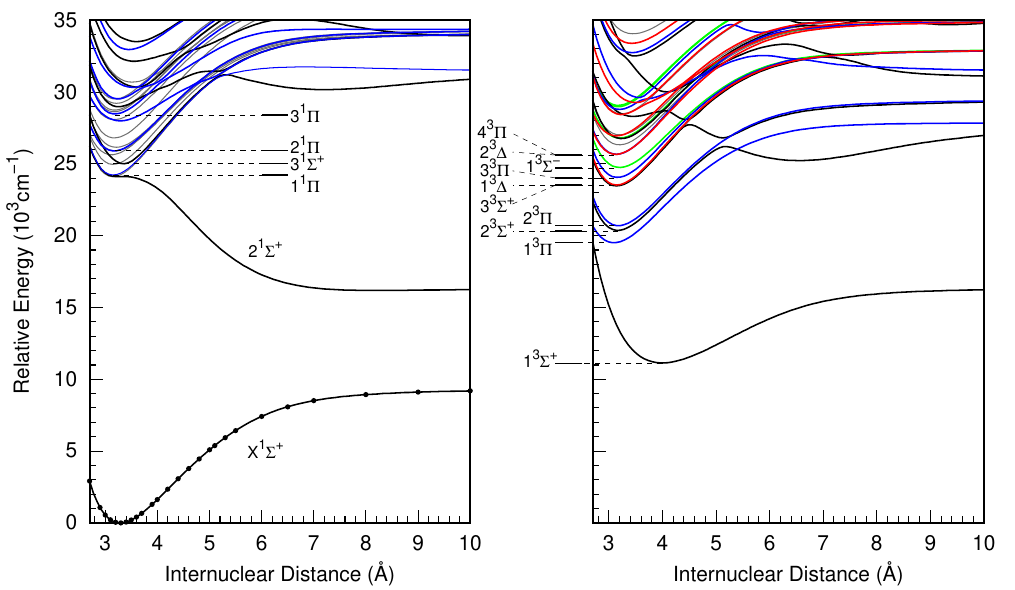}
\caption{Low-lying excited singlet (left panel) and triplet (right panel) states of YbLi$^+$ calculated with spin-free EOM-CCSD/VTZ method.
States that are able to span $\Omega=0^+$ or 1 are colored.
Symmetry $\Sigma^+$ in black, $\Sigma^-$ in green, $\Pi$ in blue and $\Delta$ in red. 
Other states are plotted in gray.
States with minima up to around 26~000~\wnumb{} are labeled.
For all calculated data points sorted per spatial and spin symmetries see SI.}
\label{fig:symmetry}
\end{figure*}

Since the neutral ytterbium and lithium atoms feature similar ionization potentials, the excited states of {YbLi}$^+$ can be attributed to two different configurations, namely Yb+Li$^+$ and Yb$^+$+Li.

We start our discussion with multi-reference methods commonly used to model excited-state structures of diatomics. 
As observed for the isolated atom, the lack of dynamic correlation in the reference wave function leads SS-RASSCF to underestimate the energies of the 4\forb[14] states compared to the 4\forb[13] ones.
However, it is computationally infeasible to (partly) recover the dynamic correlation energy via MRCISD considering the full inner-valence with 4\forb[13] configurations along the full PES.
Therefore, we calculated only single points with smaller wave function expansions to estimate the vertical singlet excitation energies of $A_1$ symmetry.
Without the inclusion of core-valence correlation and employing a minimal active space (4\forb[]-shell, $\sigma$ and $\sigma^*$), the low-lying 4\forb[] $\rightarrow$ $\sigma^*$ states are around 24~000 and 20~500~\wnumb{} for MRCISD with and without the Davidson correction, respectively.
Including the full 5\dorb[] shell in the active space shifts the vertical MRCISD transition with mainly 4\forb[] $\rightarrow$ $\sigma^*$ character to around 37~500~\wnumb{}.
Adding a Davidson correction lowers them to 19~000~\wnumb{}.
Likewise, MRCISD states featuring 4\forb[] $\rightarrow$ 5\dorb[] excitations are raised to a range between 35~000 to 40~000~\wnumb{}, which are further increased by the Q correction (between 40~000 to 50~000~\wnumb{}).
Such a large energy variation indicates an insufficient reference state in the multi-reference treatment.
Therefore, the best estimate from multi-reference methods is a small density of 4\forb[13]-type states in the range of 20~000 to 25~000~\wnumb{}, followed by a more dense region above 30~000~\wnumb{}.

As in the atomic case, EOM-CCSD allows us to target multiple distinct excitations with satisfying accuracy and relatively low computational cost.
The spin-free ($\Lambda$+S) PESs are shown in Figure~\ref{fig:symmetry}.
States dominantly composed of 4\forb[14] configurations agree with the ones found in the literature,\cite{tomza-ybli-pra-2015} namely, X$^1\Sigma^+$, 2$^1\Sigma^+$, 1$^3\Sigma^+$ and 1$^3\Pi$.
Including states mainly composed of 4\forb[13] configurations drastically increases the density of states between 20~000 and 30~000~\wnumb{}, which qualitatively agrees with the multi-reference estimate.
This region covers the 3$^3\Sigma^+$ state which directly interacts with the 2$^1\Sigma^+$ state (note the avoided crossing around 3.5 \AA{}) and the 1$^1\Pi$, 2$^3\Sigma^+$, and 2$^3\Pi$ ones.

\begin{figure*}
\includegraphics{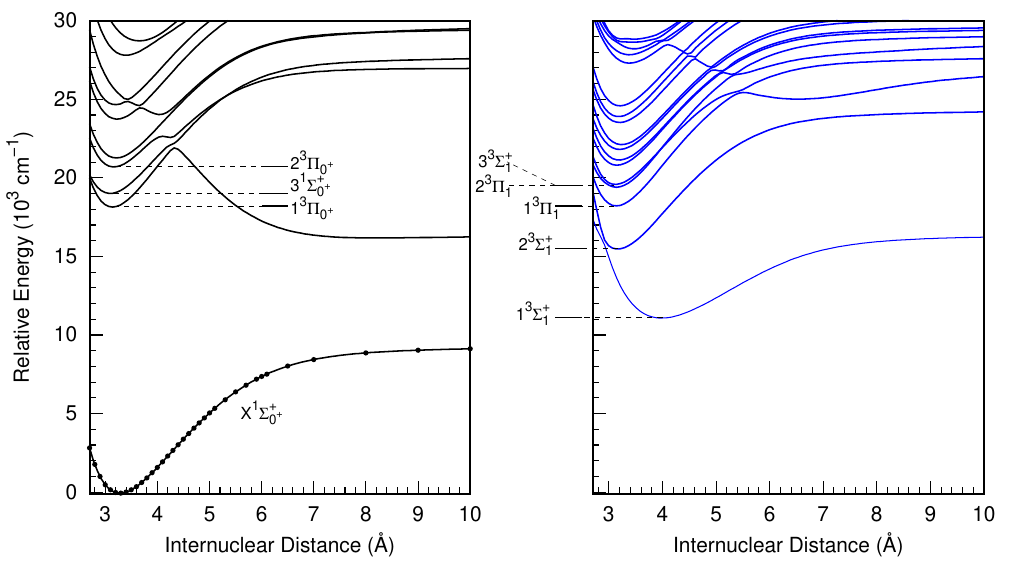}
\caption{Low-lying $\Omega$ = $0^+$ (in black; left panel)  and $\Omega$ =$1$ (in blue; right panel) states of YbLi$^+$ calculated with spin-orbit-coupling EOM-CCSD/VTZ.
The state labelling has been attributed based on the spin-free calculation (Figure~\ref{fig:symmetry}).
For all calculated data points sorted per $\Omega$ value see SI.}
\label{fig:so_pec}
\end{figure*}

Although the inclusion of states with 4\forb[13]{} character may modify the excited-state structure of the YbLi$^+$ cation at the spin-free level, these states are high in energy compared to the second dissociation channel (i.e. Yb$^+$($^2S_g$)+Li($^2S_g$)).
Therefore, all spin-free states of interest for a low energy collision of Yb$^+$+Li are mainly composed of 4\forb[14]{} configurations.
As already observed for the isolated ytterbium atom, SOC has, however, a large effect over the states of 4\forb[13]{} character, in particular, when compared to the 4\forb[14]{} counterpart.
Figure~\ref{fig:so_pec} demonstrates this change in relative energies caused by SOC deduced from the 4c EOM-CCSD results.
The PESs with minima below 21~000~\wnumb{} are labeled based on the spin-free calculation to simplify the visualization and discussion and their spectroscopic constants are summarized in Table~\ref{tab:spec_param}.
Note, however, that the SOC states feature large mixing of spin-free configurations, in particular for 1$^3\Pi_1$, 2$^3\Sigma^+_1$, and  2$^3\Pi_1$, which comprise, among othres, excitations from Yb 4\forb{} to Li 2\sorb{}, Yb 4\forb{} to Yb 7\sorb{}, Yb 4\forb{} to Yb 5\dorb{}, and Yb 6\sorb{} to Yb 6\porb{}.

Most importantly, the inclusion of SOC shrinks the relative energy gap leading to qualitative changes in the curves associated with the second dissociation channel and allowing the non-relativistic spin-forbidden triplet-singlet radioative transitions.
These changes affect the Yb$^+$+Li collision process.
First, the $\Omega=1$ state spanned by the $1^3\Sigma^+$ spin-free state mixes with the $1^3\Pi$ at small internuclear distances.
The second state of $\Omega=0^+$ symmetry suffers a much larger qualitative change, with an avoided crossing around 4.5 \AA{}.
This change in wave function character creates a meta-stable state with a relatively small potential barrier that can also affect the radiative decay of the system.

\begin{figure}
\includegraphics{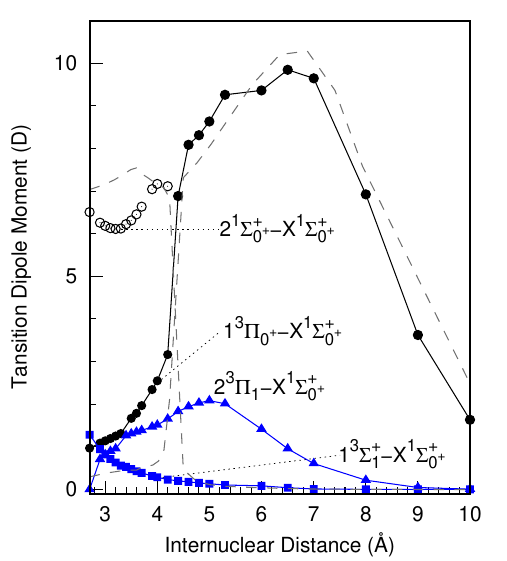}
\caption{Transition dipole moment in Debye (D) calculated between the X$^1\Sigma^+_{0^+}$ ground state and three $\Omega$ states: 1$^3\Sigma^+_1$ (filled circle), 2$^3\Pi^+_1$ (filled triangle), 1$^3\Pi_{0^+}$ (filled square) and one following the 2$^1\Sigma^+_{0^+}$ adiabatically (hollow circles). 
Values obtained using the perturbative SO-RASCI approach with unperturbed SF-EOM-CCSD energies.
Dashed gray lines marks the 2$^1\Sigma^+_{0^+}$-X$^1\Sigma^+_{0^+}$ and 1$^3\Pi_{0^+}$-X$^1\Sigma^+_{0^+}$ transition dipole moment using 4\forb[14]-configuration MRCISD+Q unperturbed energies.
}
\label{fig:so_trans}
\end{figure}

To estimate the effects of SOC and 4\forb[13]{} configurations on the light emission processes of the excited YbLi$^+$ cation, we calculate the transition dipole moments (TDMs) using the perturbative SO-RAS split with the SF-EOM-CCSD total energies as the unperturbed energies.
Figure~\ref{fig:so_trans} summarizes the TDMs with values larger than 0.5 D, between the ground state and states whose minima are lower than 22~000~\wnumb{}.
The adiabatic transition dipole values to the spin-allowed 2$^1\Sigma^+$ state (dashed line with hollow marks) agrees with the non-relativistic results found in the literature.\cite{tomza-ybli-pra-2015}
On the other hand, the diabatic curve substantially reduces the transition in the equilibrium distance (between 2.5 and 4.0 \AA{}), caused by the avoided crossing with the spin-forbidden 1$^3\Pi_{0^+}$ state.
However, the large SOC leads to non-negligible values in TDM.
Similarly, substantial TDMs are obtained for X$^1\Sigma^+_{0^+}$--2$^3\Pi^+_{1}$ and X$^1\Sigma^+_{0^+}$--1$^3\Sigma^+_{1}$, both coupled to the non-relativistic spin-forbidden transitions.

\begin{figure}
\includegraphics{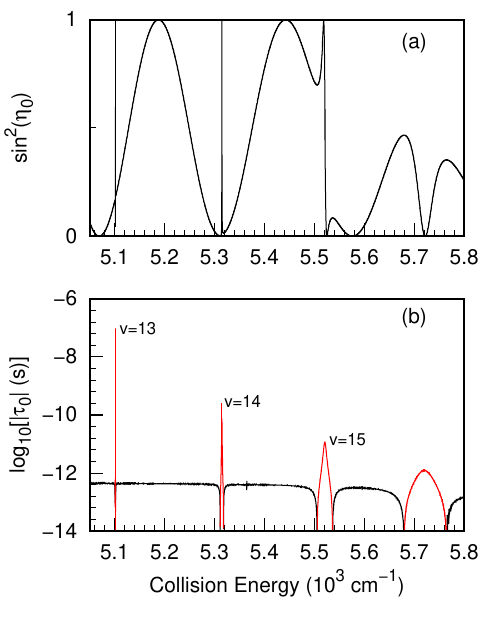}
\caption{Elastic scattering results for state 1$^3\Pi_{0^+}$ shown in Figure~\ref{fig:so_pec} of the $^{174}$Yb$^7$Li system. 
(a) sine square of the phase shift $\eta_J$, for  $J=0$.
(b) Logarithm of the absolute values of the collisional time-delay,  $\tau_J$, for  $J=0$. 
In red positive values of $\tau_0$ and in black negative ones.
The labels mark the corresponding quasi-bound vibrational states.
}
\label{fig:shift}
\end{figure}

The presence of a meta-stable (or quasi-bound) state with a non-negligible TDM to the ground state allows for an additional route via tunnelling followed by a radiative decay. 
Three resonance states are likely to be reached from tunneling through collisions, as shown via the phase shift ($\eta_J$) for $J=0$ in Figure~\ref{fig:shift}(a).
These resonances can be correlate with the vibrational levels of the meta-stable state ranging from v=13 to v=15, listed in Table~\ref{tab:vib_level}.
The pre-dissociation lifetime of these meta-stable vibrational states can be calculated by the collision delay time (Figure~\ref{fig:shift}(b)) at the resonance energies divided by a factor of four.\cite{10.1063/1.1674805}
The pre-dissociation lifetime, in particular for v=14 (0.127 ns), is comparable to the partial radiative lifetime of this state (2.472 ns).
Therefore, the tunneling route is viable and could be enhanced by stimulated emission. 
The main populated vibrational levels of the ground states should be v=10, 11, and 12, which have larger Einstein coefficients with respect to the accessible meta-stable vibrational state.
Considering misaligned collisions, that is, for $J>0$, the range of resonance energies extends from around 5~000 up to 6~000~\wnumb{}. 
Within this collision energy range between 5~000 and 5~400~\wnumb{}, the pre-dissociation lifetimes are above $10^{-10}$s.

As discussed before, it is likely that the gap between the 4\forb[14] and 4\forb[13]-type states may be underestimated in the 4c EOM-CCSD method, that is, they represent a lower limit.
An upper limit can be calculated within MRCISD+Q using only 4\forb[14] states, which is analogous to excluding any coupling between 4\forb[14] and 4\forb[13] states.
Without the presence of 4\forb[13] configurations, the 1$^3\Pi_{0^+}$ has a higher excitation energy (T$_e$=18999~\wnumb{}) and lower dissociation energy (D$_e$=$-$2823~\wnumb{}), but a similar barrier of 5~698~\wnumb{}.
As a consequence, the collisional energy required to access quasi-bound rovibrational states is not altered by the lack of 4\forb[13] configuration, but the accessible levels are $v=12$ and 13.

Additionally, with the smaller mixing among configurations the TDM is substantially reduced in the equilibrium distance (compare black and gray curves from Figure~\ref{fig:so_trans}). 
As a consequence, the partial radiative life-time of the accessible state increases slightly, namely, 37 and 27 ns for $v=11$ and 12, respectively.
The former decays mainly to the $v=8$ ground state, while the latter decays to  $v=9$, as the PESs minima are shifted from each other.
Therefore, although the presence of 4\forb[13] configurations enhances the inelastic formation of YbLi$^+$ ground state molecules, if the excited state is composed of only 4\forb[14] configurations, the tunneling route allowed via SOC remains likely to be observe experimentally.

\begin{table}[]
    \centering
        \caption{Spectroscopic parameters including  equilibrium bond length (r$_e$), intermolecular distance of ro-vibrational ground state (r$_0$), minimum electronic energy ($T_e$), vibrational constant – first term ($\omega_e$), vibrational constant – second term ($\omega_e x_e$), rotational constant in equilibrium position ($B_e$), electronic dissociation energy (D$_0$), zero-point dissociation energy (D$_e$) calculated for the $\Omega$ states of $^{174}$Yb$^{7}$Li$^+$. The r$_e$ and r$_0$ are given in Angstrom (\AA{}), while other quantities are given in wave number (cm$^{-1}$).}
    \begin{tabular}{ccccccc}
     State &  r$_e$(r$_0$)   & $T_e$ & $\omega_e$ & $\omega_e x_e$ & $B_e$ &D$_e$(D$_0$) \\
     \hline
     X$^1\Sigma^+_{0^+}$ &  3.285(3.285) &     0 & 231.13 & 1.38 & 0.232 & 9183(9068) \\
     1$^3\Sigma^+_1$     &  3.986(4.663) & 11 141 & 148.28 & 0.85 & 1.149 & 5150(4992)\\
     2$^3\Sigma^+_1$     &  3.168(3.220) & 15 531 & 260.10 & 1.00 & 0.241 & 8728(8602)\\
1$^3\Pi_{0^+}^{\dagger}$ &  3.139(3.124) & 18 203 & 259.30 & 1.64 & 0.256 &-1897(-2027) \\
     1$^3\Pi_{1}$        &  3.140(3.155) & 18 278 & 259.66 & 1.71 & 0.251 & 8222(8093) \\
     \vspace{-0.2cm}\\
     \multicolumn{7}{l}{$^\dagger$Disregarding the repulsive region between 4.3 and 10 \AA{}.}
    \end{tabular}
    \label{tab:spec_param}
\end{table}

\begin{table}[]
    \centering
    \caption{Vibrational energy levels (in \wnumb{}) for X$^1\Sigma_{0^+}^+$ and 1$^3\Pi_{0^+}$ states shown in Figure~\ref{fig:so_pec}.}
    \label{tab:vib_level}
    \begin{tabular}{ccccc}
    & \multicolumn{2}{c}{$^{174}$Yb$^{7}$Li}& \multicolumn{2}{c}{$^{174}$Yb$^{6}$Li} \\
   v &  X$^1\Sigma_{0^+}^+$ & 1$^3\Pi_{0^+}^\dagger$ &  X$^1\Sigma_{0^+}^+$ & 1$^3\Pi_{0^+}^\dagger$ \\
   \hline
   0 &    115 &   130 &  124 &  140     \\
   1 &    343 &   386 &  369 &  415     \\
   2 &    569 &   639 &  612 &  687     \\
   3 &    791 &   889 &  851 &  956     \\
   4 &   1012 &  1135 & 1087 & 1219     \\
   5 &   1229 &  1377 & 1320 & 1479     \\
   6 &   1443 &  1616 & 1550 & 1734     \\
   7 &   1655 &  1851 & 1776 & 1986     \\
   8 &   1865 &  2084 & 2000 & 2234     \\
   9 &   2071 &  2314 & 2221 & 2480     \\
  10 &   2276 &  2541 & 2439 & 2723     \\
  11 &   2477 &  2766 & 2654 & 2962     \\
  12 &   2676 &  2987 & 2865 & 3196     \\
  13 &   2872 &  3205 & 3073 & 3426     \\
  14 &   3065 &  3418 & 3276 & 3644     \\
  15 &   3255 &  3622 & 3476 & \multicolumn{1}{c}{---}     \\
     \vspace{-0.2cm}\\
     \multicolumn{5}{l}{$^\dagger$Disregarding the repulsive}\\
     \multicolumn{5}{l}{region between 4.3 and 10 \AA{}.}
    \end{tabular}
\end{table}

\section{Conclusions}
This work provides a fully relativistic \textit{ab inito} picture of the ground- and excited-state of Yb, Yb$^+$ and \ce{YbLi+} based on the 4c Dirac--Coulomb Hamiltonian and the EOM-CCSD approach.
For the first time, we report a complete electronic spectrum of the Yb atom, which includes all possible single $4f\rightarrow 5d$ electronic transitions.
Our theoretical data agrees very well with the NIST reference atomic levels in the lower part of the spectrum and complements the missing $4f\rightarrow 5d$ block of 10 transitions in the Yb spectrum. 
Due to the incompleteness of the available basis sets and the limited amount of orbital relaxation effects in our theoretical model, the agreement with the experimental Yb spectra of the $4f\rightarrow 5d$ block is less satisfactory. 

Nonetheless, we show that the impact of the  $4f\rightarrow 5d$ excitations on the electronic spectra of \ce{YbLi+} is nonnegligible, especially when spin-orbit coupling is accounted for from the beginning in the 4c formalism, which results in the lowering of the $4f\rightarrow 5d$ block of transitions by a few thousand cm$^{-1}$.
That, in turn, affects the resulting \ce{YbLi+} spectroscopic characterization.
Specifically, we observe that:
\begin{enumerate}
\item[$\bullet$] the second $\Omega=0^+$ state is no longer repulsive and forms a meta-stable state
\item[$\bullet$] that meta-stable state leads to quasi-bound rovibrational levels that are accessible via tunneling
\item[$\bullet$] the presence of \forb[14]-\forb[13] interactions increase the transition dipole moment and, as a consequence, the probability of radiative-decay to the ground-state of \ce{YbLi+}. 
\end{enumerate}

In conclusion, our results suggest that higher collision energies (above 5~000~\wnumb{}) should be used in experimental studies to increase the formation rates of the bound ground-state \ce{YbLi+} molecules. 

\begin{acknowledgments}
M.G.~acknowledges financial support from The Ulam Programme -- Seal of Excellence grant of the Polish National Agency for Academic Exchange, Poland (Grant No.~BPN/SEL/2021/1/00005/U/00001).
P.T.~acknowledges financial support from the OPUS research grant from the National Science Centre, Poland (Grant No.~2019/33/B/ST4/02114).
We acknowledge that the results of this research have been achieved using the DECI resource Bem (Grant No.~412) based in Poland at Wroclaw Centre for Networking and Supercomputing (WCSS, http://wcss.pl) with support from the PRACE aisbl. 
Funded/Co-funded by the European Union (ERC, DRESSED-pCCD, 101077420).
Views and opinions expressed are, however, those of the author(s) only and do not necessarily reflect those of the European Union or the European Research Council. Neither the European Union nor the granting authority can be held responsible for them. 
\end{acknowledgments}
\bibliography{rsc.bib}

\newpage
\centering
\includegraphics[]{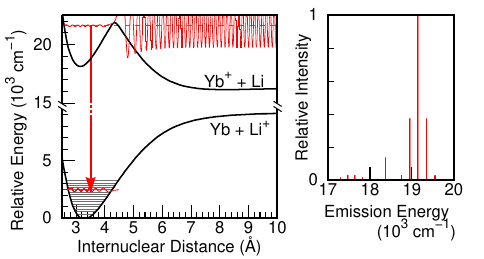}
\end{document}